\documentclass[preprint2]{aastex}

\def\etal{\mbox{et al.\/}}

\def\m{$^{\rm m}$}

\def\lae{\mathrel{<\kern-1.0em\lower0.9ex\hbox{$\sim$}}}
\def\gae{\mathrel{>\kern-1.0em\lower0.9ex\hbox{$\sim$}}}
\def\m{Mrk~421\ }

\def\euve{\mbox{\em EUVE\/}}
\def\asca{\mbox{\em ASCA\/}}

\slugcomment{Submitted to {\it The Astrophysical Journal} 1 November 1999}

\shorttitle{Cagnoni \& Fruscione}
\shortauthors{Extreme Ultraviolet Spectra of \m}

\begin{document}

\title{Four Years of Extreme Ultraviolet Observations of Markarian 421.
I: Spectral Analysis}
\author{I. Cagnoni\altaffilmark{1,2} \and  A. Fruscione\altaffilmark{2}}
\affil{$^{1}$ SISSA, Via Beirut 4 -34138, Trieste, Italy}
\affil{$^{2}$ Harvard-Smithsonian Center for Astrophysics, 60 Garden Street, 
Cambridge, MA 02138, USA}
\email{ilale@sissa.it}

\begin{abstract}

We analyzed the $\sim 950$ ks of spectroscopic data accumulated by the
{\it Extreme Ultraviolet Explorer} ({\it EUVE}) satellite between 1994
and 1997 for the BL Lacertae object Markarian 421.  The EUV spectrum
is well detected in the 70--110 \AA\ (112--177 eV) range and can be
fitted by a power law model plus 
an absorption feature in the $\sim71-75$ \AA\ range.
Previous studies
of EUV absorption features in \m and in the other EUV bright BL Lac
object, PKS2155$-$304, explain this absorption feature
 as a superposition of Doppler-smeared absorption lines
(mainly L- and M-shell transitions of Mg and Ne) originating in
high-velocity gas clouds ionized by the beamed continuum of the
associated relativistic jet. We show that, for example, Fe IX L could
also be a possibility consistent with the marginal detection of oxygen
absorption lines in the X-ray range. However physical models are highly
sensitive to the assumptions on the photoionizing continuum and the
surrounding gas.

\end{abstract}


\section{Introduction}

Evidence for absorption features in the X-ray spectra  of BL Lac objects at
$\sim 0.6$ keV \footnote{PKS~2155-304 (Canizares \& Kruper 1984), H1426+428
 (Sambruna et al. 1997) and PKS~0548-322 (Sambruna \& Mushotzky 1998)}
 has existed for many years.
This is a tantalizing situation: BL Lac objects were originally defined 
by their lack of the sharp spectroscopic features that provide the normal
tools for understanding astrophyisical objects.
Without this tool, only large multisatellite, multitelescope campaigns
 spanning 20 decades of frequency have been able to constrain the BL Lac 
emission mechanism.
Absorption features can provide a unique tool to study the presence and 
the dynamical properties of any gas surrounding the nucleus of the otherwise
 `featureless' BL Lac objects.
The extremely large Extreme Ultraviolet Explorer ({\it EUVE}) satellite database on \m offers a so far a unique 
opportunity to search for and characterize any sharp spectroscopic feature.

\m is one of the brightest, best known and closest BL Lac objects ($z$=0.0308,
Ulrich et al.  1975).
\m was in fact the first BL Lac object found to have X-ray emission (Richetts et al. 1976) and its low redshift makes 
this object  one of the few extragalactic sources  detected at $\gamma$-rays
above 300 GeV up to TeV energies (Punch et al. 1992).
\m has been
extensively observed at radio (e.g. Zhang \& Baath 1990),
UV/optical (Maza, Martin \& Angel 1978; Mufson et al., 1990) and X-ray
frequencies (e.g. Mushotzky et al., 1979; George,
Warwick \& Bromage 1988, Takahashi \etal\ 1996, Guainazzi \etal\ 1999).
\m shows optical polarization, a flat radio spectrum and significant
time variability, characteristics of the blazar class (Makino et al.,
1987) and its  parent galaxy has been identified as a
giant elliptical (Ulrich et al. 1975; Mufson, Hutter \& Kondo 1989).

In the EUV band ($\sim 60-750$ \AA) \m was detected during the {\it
ROSAT} WFC survey (Pounds et al. 1993, Pye et al. 1995) and the {\it
EUVE} all-sky survey (Marshall, Fruscione \& Carone 1995, Fruscione
1996).  It is one of the strongest EUV extragalactic sources and was
immediately recognized as a prime target for subsequent EUV
spectroscopic studies.  For 4 almost consecutive years (1994, 1995,
1996 and 1998) \m was the target of multiwavelength campaigns covering
radio to TeV energies (Macomb et al., 1995, Kerrick et al. 1995,
Buckley et al. 1996, Takahashi et al. 1999) and the {\it EUVE}
satellite (which gives simultaneous photometric and spectroscopic
data) was always part of the campaigns.  In addition, in 1995 and in
1997 \m was observed serendipitously with the {\it EUVE} photometers
(the ``scanners'').\\ The large amount of data accumulated for \m by
the {\it EUVE} satellite to date ($\sim 1100$ ks of public imaging
data and $\sim 950$ ks of public spectroscopic data) represents the
best coverage for any BL Lac object at high energies.  Less than half
of this data has been analysed to date (Fruscione et al. 1996, Kartje
et al. 1997, hereafter K97) revealing that (i) the source exhibits a
flare-like behavior (ii) the strong EUV variability is correlated with
soft X-ray and TeV-energy emission, and (iii) the spectrum may have
strong absorption features between 65 \AA\ and 75 \AA\ (0.16-0.19 keV)
(K97).  However in the EUV, all the previous analysis of BL Lacs can
only infer the presence of absorption from the extrapolation of the
X-ray spectrum (e.g. Konigl et al 1995 for PKS2155-304 and K97 for
MRK~421).

The purpose of this paper is to analyze, in an homogeneous way, all of the
950 ks of {\it EUVE} public spectroscopic data for \m. 
The results of this paper will also provide a useful reference for the
calibration of other satellites observing in this energy range.  In
particular \m is a calibration target for both the Chandra and XMM
satellites, whose Low Energy Transmission Grating (LETG) and
Reflection Grating Spectrometer (RGS) partially overlap with the {\it
EUVE} energy band.\\
In Section 2 we present all the {\it EUVE} spectra of \m taken
from 1994 to 1997; in Section 3 we discuss the results and in Section
4 we give a brief summary and our conclusions.  All the photometric
data, the corresponding lightcurves and a detailed variability
analysis are presented in a companion paper (Cagnoni, Papadakis \&
Fruscione, 2000, hereafter paper {\sc ii}).

\section{Observations}

The {\it EUVE} satellite was launched in
June 1992 and is still acquiring astronomical data in the wavelength
range from $\sim 60$ to 750 \AA.  Onboard {\it EUVE} there are 4
telescopes: three of them are co-aligned photometers observing in 4
different EUV bandpasses, while the fourth is the Deep
Survey/Spectrometer (DS/S, see e.g. Welsh et al. 1990) mounted
orthogonally to the scanners. (Figure~1 of paper {\sc ii} 
shows the DS effective area compared to the scanner effective area).
In addition to a photometer (10 \% bandpass 68-178 \AA) the DS/S
telescope is also equipped with three spectrometers (Hettrick \&
Bowyer, 1983; Abbott et al. 1996) covering the ``short'' (SW:
$70-190$ \AA), ``medium'' (MW: $140-380$
\AA) and ``long'' (LW: $280-760$\AA) EUV wavelengths.  This
configuration allows simultaneous imaging and spectroscopy with a
spatial resolution of $\sim 1^{\prime}$ and a spectral resolution of
$\lambda/\Delta
\lambda \sim 200$ at the short wavelengths, i.e. 0.35 \AA \/ at 70 \AA .\\
Further details on the science instrumentation and performances can be
found in Malina (1994), Sirk et al. (1997) and Abbott et al. (1996).

The scanners were used to carry out the {\it EUVE} all-sky survey
(July 21 1992 - January 23 1993) and are currently utilized in the
Right Angle Program (RAP), that is the observation of targets of
scientific interest with the scanner telescopes while the DS/S
telescope is conducting the primary science observation.  The DS/S has
been used for a deep EUV survey along the ecliptic during the all-sky
survey and for pointed spectroscopic and imaging observations since
then.

We use throughout this paper all of the publicly available DS/S data on
MRK~421: from 1994 to 1997, a total of 950 ks.  We are not considering the most recent
observation (April 1998) since the data is presently being analysed by the
original proposers (Marshall \etal\ in preparation). 
\m was observed by {\it EUVE} several times from
1994 to 1997, four times with the DS/S and twice with the scanners;
Table~1 summarizes the observations.\\ Because of strong
absorption by the interstellar medium (ISM) along the line of sight
(even at the low $N_H=1.45 \times 10^{20}$~cm$^{-2}$ Elvis, Wilkes \& Lockman 1989),
\m was detected only at the shortest wavelengths: in the Lex/B filter
of the scanners (10\% bandpass 58-174~\AA ), in the Lex/B filter of
the DS instrument (10\% bandpass 67-178~\AA) and, relevant for this paper,
 in the SW spectrometer (70-190 \AA)
\footnote{A graphical representation of the simulated DS/S and scanner
responses is presented in Figure 2 of paper {\sc ii}, where we
convolved the \m energy spectrum corrected for the ISM absorption with the
DS/S and scanners effective area.  No signal is present at wavelengths
longer than $\sim 110$~\AA}.

\subsection{Energy spectra}

After rejecting the time intervals with high particle background, such
as the satellite passages over the South Atlantic Anomaly, and
correcting the data for instrumental deadtime and telemetry
saturation, we measured a total effective exposure of 315858~s,
506780~s, and 354806~s for the 1994, 1995 and 1996 observations
respectively (note that the first 1995 and  the 1997 observations
 were performed with the scanners, instrument with imaging
capabilities only).
\\
We extracted all the spectra in a homogeneous way from the
two-dimensional SW detector image, using the {\it EUVE} Guest Observer
Center software (IRAF/EUV package) and other standard spectroscopic
IRAF tasks.  The spectrum was extracted in a 14 pixel wide aperture
and the background obtained by linearly fitting an average background
in two regions ($\sim 80$ pixel wide) of the detector, one on each
side of the spectrum.  The resulting wavelength-calibrated spectra
(Figures 1, 4 and 8) were convolved with the proper effective area and
binned over 1 \AA.  For the 1994 on-axis observation (Figure~1) we
used the on-axis effective area derived from in-orbit calibrations for the 75 \AA - 110 \AA \/ range.
The 1995 and 1996 observations were
performed off-axis (a technique that slightly increases the
short-wavelength coverage, down to $\sim 70$~\AA ).  This results in a
different effective area (Marshall et al. 1999) that we applied here
for the first time.  The 1$\sigma$ error bars on the flux measurements shown
in the figures of all the spectra were computed using the formula
\[\sigma(\lambda)=\frac{\sqrt{[S_{\lambda} + 
B_{\lambda}/(1+H_{sp}/H_{B})]}}{A_{eff}(\lambda)} \]
where S is the signal from the target, B the averaged background, both in counts/s,
  $A_{eff}$ the SW effective area in $cm^2$  and $H_{sp}$ and $H_{B}$ are
respectively the width in pixels of the extraction region
for spectrum (14 pixels)  and background (160 pixels) in the
direction perpendicular to the dispersion.

We modeled the EUV spectrum of \m using version 2.0 of the SHERPA
modelling and fitting software
(recently developed at the Chandra X-ray Center). 
We initially used an absorbed power law of the form

$$\rm{f}(\lambda)=\rm{f}(\lambda_0)\lambda^{\alpha-1}exp\{-[\Sigma_{\rm{X}}
\rm{N(X)} {\sigma}_X)\} ~~~~~\rm{photons\ cm}^{-2}\ \rm{s}^{-2}\  \AA^{-1}\/ $$


where f($\lambda_0$) is a normalization factor.
We used $\lambda_0=80$ \AA\  (E$_0$=0.155~keV). 
$\alpha$ is the energy index and
the absorption is characterized by  a column density N(X) and an
absorption cross section $\sigma_{\rm{X}}$ for each element.
We included in the model H, He {\sc i} and He {\sc ii} (Rumph, Bowyer
and Vennes, 1994) and heavier elements (Morrison and McCammon, 1983).  
We fixed the Galactic hydrogen column density at $N_H =1.45 \times
10^{20}$~cm$^{-2}$ (Elvis, Wilkes  \& Lockman  1989) and the ratios
$N_{He{\sc i}}$/$N_{H{\sc i}}$=0.1  and $N_{He{\sc ii}}$/$N_{H{\sc i}}$=0.01. 
The best fit values for the simple absorbed power law model are
summarized in Table~2 together with their $1 \sigma$ uncertainties for 1
interesting parameter.
Two-parameter ${\chi}^2$ contour confidence levels at
${\chi}^2_{min}$+1, +4.61 and +5.99 -- corresponding respectively to
the 1$\sigma$ uncertainty for each single parameter and to the 90\%
and 95\% joint confidence levels -- are shown in Figure~2, 5 and 9 for 
the 1994, 1995 and 1996 data sets respectively.

\subsubsection{1994 energy spectrum}

Our best fit flux in 1994 is slightly higher than the previous 
analysis of Fruscione et al., 1996 
(f$_{\rm 80\AA} = 6.44\pm 0.33 \times 10^{-3}$ photons
 cm$^{-2}$ s$^{-1}$ \AA$^{-1}$ compared to their f$_{\rm 80\AA} = 
290 \pm 15$ $\mu$Jy$= 5.47\pm0.28\times 10^{-3}$ photons cm$^{-2}$ s$^{-1}$ 
\AA$^{-1}$) but our slope is within the errors ($\alpha = 
0.85^{+0.66}_{-0.72}$ compared to their $\alpha =1.4 \pm 0.8$).
The 1$\sigma$ error
bars reported in Fruscione et al., 1996 were not corrected for the
effective area; the small difference in the value of the normalization
can be possibly explained by the difference in the errors and a
possible difference in the wavelength range used to model the
spectrum. This is the only spectrum for which the addition of an
absorbed interved gaussian line does not improve the fit (Tab.3).

\subsubsection{1995 energy spectrum}

The 1995  \euve\ DS/S observation was part of a large multiwavelength campaign
during which a powerful TeV flare occured (Buckley \etal\
1996). Simultaneously an X-ray flare developed (Takahashi \etal\
1996), and the EUV followed with a short delay (Buckley \etal\
1996). As illustrated in figure 3,  the
\euve\ observation spanned the flare ($\sim 104$ ks), the decay
($\sim 218$ ks) and the following quiescent state ($\sim 185$ ks)  
The observation was performed $\sim 0.3^{\circ}$
off-axis and as expected the signal reaches down $\sim70$\AA.  This
spectral region is crucial since it covers the location of the wide
EUV absorption feature proposed by K97 using less than half (the flare decay) 
 of the 1995  data.  In our analysis we
corrected the data with the  $\sim 0.3^{\circ}$ off-axis
effective area (Marshall \etal\ 1999) and the resulting spectrum,
binned over 1 \AA, is presented in Figure~4. 

\noindent In order to compare our data analysis to that of
 K97, and in an attempt to look for changes during
the flare, we divided the 1995 observation into three parts (Figure~3):\\
\noindent Flare: April 25--28 ($\sim 104$ks), the time interval of the \euve\ flare;\\
\noindent Decay: April 29 -- May 6 ($\sim 218$ ks), the time interval analysed by
K97;\\
\noindent Quiescence: May 7--13 ($\sim 185$ ks), the remaining data.
The {\it EUVE} spectra for the 3 parts are shown in Figiure~6 a,b and c.

Table 1 lists the exposures times and Table 2 summarizes the best-fit
parameters for each one of the three parts.  In Figure 7 we plot the
two-parameter ${\chi}^2$ contour confidence levels.  

\subsubsection{Comparison with previous results}

The `decay' section of the 1995 spectrum deserves comparison
with the analysis of K97, given the importance of the absorption line evident in
their Figure~2.
We show the  effect of using the improper $\sim 0.3^{\circ}$ off-axis  
effective area in Figure 6b; we compared  the \euve\ spectrum obtained using the
on-axis effective area (open cirles) as in K97,
and the $0.3^{\circ}$ off-axis effective area (filled squares). 
The effect of using the appropriate calibration 
is to increase the flux below $\sim 80$ \AA\
and so weaken the absorption feature ($\sim 45$\%). 
Another effect to consider is that K97
 evaluated the level of the continuum in the EUV range by 
adopting an $\alpha_{\rm EUVX}$ ($\approx1.32$) slope derived from
the 0.15 kev EUV flux and the 1.5 keV X-ray (\asca) flux, which
resulted in a strong absorption feature (see also \S 3 for
further discussion on this point).

Our best fit slope $\alpha = 2.19 \pm 0.5$  (Table~2)
 is flatter than the $\alpha = 3.5 \pm 0.8$ in K97.
If we consider the same wavelength range (70-95 \AA\ instead of 70-110 \AA) 
we find $\alpha = 2.54 \pm 0.56$, consistent with K97 result.

A simultaneous fit of an absorbed  power law plus an absorbed (inverted)
 gaussian should provide a more reliable result.
If the position of the
gaussian is a free parameter then the reduced $\chi ^2$ goes to
1.17 and a narrow (FWHM=$1.8 \pm 0.5$ \AA) absorption feature is found at 
$\sim 72.7 \pm 0.3$ \AA\ while the slope becomes significanlty flatter 
($\alpha=1.62 \pm 0.59$) (see Tab.3).
This is the first time that an absorption feature in the EUV range is 
derived from EUV data alone. 

Both for PKS~2155-304 (Konigl \etal\ 1995) and for \m (K97) the
presence of absorption features was previously derived on the basis of an
extrapolation of the X-ray  power law and not directly from {\it
EUVE} data. 
However there is strong evidence that the X-ray spectrum of the
high-energy peaked BL Lac objects, including \m, is concave (Sambruna et
al. 1994, 1997; Tashiro et al. 1995; Takahashi et al. 1996a; Giommi et
al. 1998; Guainazzi et al. 1999) and a gradual steepening with energy
agrees with the ``synchrotron self Compton'' scenario, which has often
been used to explain the spectral energy distribution of BL Lac
objects (Ghisellini, Maraschi \& Treves 1985, Ghisellini 1989).
Therefore, a simple power law model is an unreliable
extrapolation from  X-rays to the EUV.

In order to verify this assumption for \m, we reanalyze, as an
example, one of the many one-hour-long 1995 {\it ASCA} observations
(the one on 25-26 April) undertaken simultaneously with the {\it EUVE}
observation (1995 `flare'). This in order to investigate the spectral slope
below 1.5 keV (Takahashi \etal\ 1996b and 1999 only published data for
1.5 keV and above).  First we fitted the \asca\ spectra with an
absorbed broken power law and the best fit requires a break position
at $E_{break}=1.77 \pm 0.07$~keV and the energy slopes varies from
$1.44 \pm 0.16$ above the break to $0.98 \pm 0.02$ below the break
(consistent with what found by Takahashi et al.).  The broken power
law model however does not give a good fit of the spectrum below 1.5
keV; a better result is found if we assume a ``curved spectrum''
parametrized as
\[F(E)= E^{-(f(E) \alpha _{low} + (1-f(E)) \alpha_{high})}\]
where $f(E)=(1- exp(-E/E_0))^{\beta}$, $\alpha _{low}$ and
$\alpha_{high}$ are the low and high energy asymptotic slopes
and $\beta$ is the curvature radius in the energy space.
The best fit values for the curved  model with fixed $\beta =1.0$ 
(e.g. Guainazzi \etal\ 1999) are 
$\alpha_{low}=0.57 \pm 0.06$, $\alpha_{high} = 1.34 \pm 0.13$ and
$E_0=1.78 \pm 0.18$ (see fig. 10); 
these values show that the slope flattens toward
lower energies and that an extrapolation with a simple power law is
not an accurate way to predict the EUV continuum level; the low 
energy asymptotic slope ($\alpha_{low}=0.57$) is generally in agreement
with the flat slope derived from the EUV spectrum
($\alpha=0.71\pm0.61$ from Tab.3).
 \\
Note that in the case of PKS~2155-304 there is an additional problem: 
the {\it ROSAT} spectral slope used for the extrapolation was not measured
simulteneously to the EUV spectrum and the spectral slope has been
observed to harden when the source intensity increases (Giommi et
al. 1998).

\subsubsection{1996 energy spectrum}

In 1996 \m was again observed $\sim 0.3^{\circ}$ off-axis with a total
effective exposure of 354806~s.  The 1 \AA\ binned spectrum and the
best fit parameters model are plotted in Figure~8; the corresponding
contour confidence levels are shown in Figure~9.

\subsubsection{The EUV absorption feature}

We performed several fits
using an absorbed power law model plus gaussian for all the
observations. The results are listed in Table~3.
In every case, except for the 1994 observation,
the addition of a gaussian absorption line improves the
fit. 
The power law slope is in the range $\alpha=[-0.15,1.62]$ 
which is not too distant from the slope measured 
from X-ray data, (see e.g. Guainazzi \etal with SAX data which gives 
$\alpha=0.3-1.1$ depending on the spectral model).
The best-fit center of the gaussian absorption lines is in the 71-75
\AA range with full width half maximum varying from 0.2 to 6 \AA.
We note however that the 6 \AA\ FWHM is relative to the global 1995
observation and it is probably due to the superposition of
narrower slightly variable lines.

%


\section{Discussion}

\subsection{The absorption feature}

Ever since {\it Einstein} Pioneering observations suggested the
existence of an ubiquitous absorption feature around $\sim 0.55$ keV
in the X-ray spectra of BL Lacertae objects (Canizares \& Kruper 1984, 
Madejski et al. 1991),  much X-ray and EUV data has been collected with the goal of
confirming the existence of such absorption features. 

\subsubsection{EUV/X-ray absorption in other BL Lac objects}

The first object to show evidence of an X-ray absorption feature was
PKS~2155-304 ($z =0.117$) in a 1980 {\it Einstein} grating spectrum 
(Canizares \& Kruper 1984). The feature was detected at $\sim 0.6$ keV
 ($\lambda = 21$ \AA) and extended for 50-100 eV ($\lambda = 120-250$ \AA); 
it was interpreted by the authors as O~VIII
Ly$\alpha$ emitted by high velocity material.  Looking at a 1979 {\it
Einstein} solid state spectrometer observation of the same object Madejski 
et al. 1991 found evidence for absorption at $\sim 0.55$ keV, 
consistent with the grating result.\\
 Krolik et al. 1985 modelled the absorption as due to an outflowing 
subrelativistic wind beamed in our direction.\\ PKS~2155-304 was
subsequently observed by {\it BBXRT} in 1990 and the absorption
feature at $\sim 0.55$ keV was confirmed in those data (Madejski et
al. 1994), although  instrumental effects cast some doubt on this result 
(Sambruna \etal\ 1997, Weaver \etal\ 1995).\\ 
In a long exposure  ($> 100$ ks) {\it Beppo-SAX} spectrum of PKS~2155-304
in the soft X-rays (Giommi et al. 1998) the best fit  model is improved 
(at 99\% confidence) when a notch at $0.55 \pm 0.04$ keV is added to the model,
although the magnitude of the feature is close to the instrument calibration limits. \\
PKS~2155-304 was also observed  by {\it EUVE}. A 1992  30 ks
observation gives  hints of an absorption feature at $\sim 80$ \AA\ (Fruscione et al. 1994).
 Two longer observations in June (130 ks) and in July (150 ks) 1993 
 suggested the existence of 2 to 5 absorption features at a $\ge 3
\sigma$ level in the range 75 to 110 \AA \/ (Konigl et al. 1995).
Extrapolating the {\it ROSAT} X-ray spectral slope into the EUV the
authors found evidence of a wide absorption feature from $\sim 75$ to
$\sim 85$ \AA\/ and interpreted it as the superposition of Doppler-smeared
absorption lines (L-shell transitions of Mg and Ne as well as M-shell
transitions of Fe) that originate in a high-velocity clumped outflow
from the nucleus.  According to this model the
X-ray line spectrum should be dominated by an O~VII K$\alpha$
broadened and blueshifted feature. This corresponds to the feature
observed in by {\it Einstein} and in 1990 by {\it BBXRT},
 but the predictions are an order of
magnitude greater than the observations; to match {\it BBXRT}
measurement the oxygen must be present in subcosmic abundance (Konigl
\etal\ 1995).

Another BL Lac object to show an X-ray absorption feature is
H1426$+$428 ($z=0.129$). Sambruna et al. (1997) reanalyzed a {\it BBXRT} 1990
observation  and found a broad feature around $0.5 - 0.6$ keV confirmed by 
an {\it ASCA} 1994 spectrum, but not by a {\it ROSAT} 1993 spectrum.

A third BL Lac, PKS~0548-322 ($z=0.069$) (Sambruna \& Mushotzky, 1998) 
confirmed the
early discovery of an absorption feature from the {\it Einstein} spectrum
(Madejski et al. 1991); this feature is present in {\it ASCA} data and
can be modeled with either an edge at $\sim 0.66$ keV and optical
depth $\tau \sim 0.3$ or with a notch at $\sim 0.82$ keV and fixed width
of $\Delta E = 0.1$ keV and covering fraction $f_c \sim 0.1$.
The {\it ASCA} absorpton feature is at energies significantly higher
than in previous X-ray observations, when the continuum was in a
slightly higher state (by a factor $\sim 1.4$); this provides tentative
evidence for variability in the absorption feature.

\subsubsection{EUV/X-ray absorption in Markarian 421}

K97 interpret the absorption  feature in the context of the Konigl
\etal\ (1995) model for PKS~2155-304 as a superposition of Mg~VIII and
Mg~IX absorption lines. 
The implied ionization state of the gas would require stronger O~VII 
absorption in the X-ray range than is allowed by observations
and the authors invoke subcosmic oxygen abundance to explain its absence.

Our analysis of the 1994, 1995 and 1996 {\it EUVE} spectra of \m shows
an absorption feature inferred directly from the \euve\ data,
(i.e. by modelling the EUV continuum and the absorption line 
simultaneously).  The simplest model for this feature is a reverted
Gaussian at about 71-75 \AA \/ a maximum FWHM of the order of few \AA.  


In order to investigate likely absorption features in a BL Lac environment,
we ran photoionization equilibrium models (Nicastro \etal\ 1999, 
Nicastro \etal\ 2000) which include the strongest 300 emission and resonant 
absorption lines down to 50 eV  (oscillator strength $>0.1$).  We find
that the predicted wavelengths and strengths of the absorption lines,  
both  absolute and relative, are highly sensitive to the properties
of the photoionizing continuum and the absorbing gas (e.g. the
ionization state and the assumed outflowing velocity). 
The number of transitions in the model also has a strong effect.
However assuming the observed EUV-to-X-ray continuum and adjusting the
gas paramters to maximize the relative abundance of OVII ($\sim65\%$)
both an OVII K$\alpha$ absorption line (EW=0.25\AA) and a FeIX L
($\lambda_{\rm rest}=82.43$ \AA, EW=0.2\AA) absorption line are expected
respectively around 0.6 keV and 73 \AA\ (assuming an outflow velocity
of 0.15c). The equivalent width of the lines is too small 
to give significant detections, however, especially in the EUV
regime, the presence of many fainter satellite lines (not included in
the model) could explain the observed feature
For example between the Si VI line 
($\lambda=72.34$ \AA) and the Mg VI complex
($\lambda~ 71.3$ \AA), both included in the model, at
least 15 additional lines from ionized Fe, Al, Be and Ne are not
included (oscillator strength 0.01-0.1) and could be important especially
if the elements are abundant.

Even though no strong absorption line seem to be visible in the X-ray
spectrum, we note that both in the  1979 {\it Einstein} data (Madejski et al. 1991) and in the  1997 and 1998 {\it Beppo-SAX} spectrum
(Guainazzi et al. 1999, Fossati et al. 2000) the fit improves when adding an absorption
feature centered at $\sim 0.55$ keV.  



In order to investigate any possible correlation between spectral changes
and source flux, we plotted in Fig.11 the energy index as a function
of total EUV flux for the the power law plus gaussian models. 
However the large error bars prevent any statististically significant 
conclusion. 

\section{Conclusions}

From {\it EUVE} data, we have found a clear absorption feature at
$\sim 71-75$ \AA\ in the spectrum of MRK~421.  The absorption features
detected in the EUV/X-ray spectra of BL Lac objects are often poorly
determined statistically, as in the {\it Beppo-SAX} observations of
PKS~2155-304 and MRK~421 (Giommi et al. 1998 and Guainazzi et
al. 1999), but it is interesting to notice that different instruments
on different satellites point to the same possible $~0.55$ keV feature
and to a (probably linked) absorption feature in the EUV spectra of
the same objects.

We do not know if X-rays and EUV absorption features are present in
all or many BL Lac objects, as initially suggested by Madejski et
al. 1991, but to-date absorption features at $0.5-0.6$ keV have been
found -- or at least statistically improves the spectral fitting -- in
four of the brightest BL Lacs objects (PKS~2155-304, H1426$+$428,
PKS~0548-322 and MRK~421) and EUV absorption features has been found
in two of them (PKS~2155-304 and MRK~421).  Future observations
covering the EUV to X-ray energy bands with much higher resolution and
collecting area, as with {\it Chandra} and {\it XMM}, will soon be
able to answer this question and give more detailed observational data
on which to develop new models.

\acknowledgments
We would like to thank Dr. Herman Marshall for providing the
off-axis effective areas for the \euve\ DS instrument, and
Drs. I.~Papadakis, A.~Celotti, R.~Sambruna, M.~Elvis for the useful
discussions.  We are particularly grateful to F.~Nicastro and G.~Matt
for making their models available to us.  We thank the \euve\ science team
and in particular Dr. Roger Malina for their logistical support for
our continuing analysis effort on \euve\ data.  This research has made
extensive use of the High Energy Astrophysics Science Archive Research
Center Online Service (HEASARC), provided by the NASA-Goddard Space
Flight Center, of the NASA/IPAC Extragalactic Database (NED) operated
by the Jet Propulsion Laboratory, Caltech, under contract with NASA
and of NASA's Astrophysics Data System Abstract Service.  A.F. and
I.C. are grateful to the Brera Observatory for the ospitality.  This
work was supported by AXAF Science Center NASA contract NAS 8-39073
and by NASA grants NAG 5-3174 and NAG 5-3191.

\clearpage

\figcaption{EUV spectrum of MRK~421 obtained in 1994 with the SW
spectrometer.  The best fit power law model is overplotted (solid line).
The lower panel shows the residuals to the fit.\label{fig1}}

\figcaption {${\chi}^2$ contour confidence levels 
for normalization and energy index of the 1994 spectrum.  The contours
correspond to ${\chi}^2_{min}$+1, +4.61 and +5.99, respectively the
1$\sigma$ uncertainty for each single parameter and the 90\% and 95\%
joint confidence levels. The best fit is marked by the filled
symbol.\label{fig2}}
 
\figcaption{April 25 - May 13 1995 
Deep Survey lightcurve binned over one average
{\it EUVE} orbit ($\sim 5544$ s). The ``flare'', ``decay'' and
``quiescence'' intervals are marked.\label{fig3}}

\figcaption{EUV spectrum of MRK~421 obtained in 1995 with the SW
spectrometer during a $0.3^{\circ}$ off-axis observation.  Hence note
the increased coverage toward shorter wavelengths.  
The best fit power law model (dotted line) and the best fit power law +
inverted gaussian model (solid line) are shown together with the
residuals to the fits.
\label{fig4}}
 
\figcaption{Same as Fig.~2 for the 1995 spectrum and the power law model.\label{fig5}}

\figcaption{Individual sections of the 1995 spectrum of MRK~421
(a) During the flare (MJD 49832-49835 or Apr. 25-28); (b) during April
29 - May 6(MJD 49835-49843) for direct comparison with the analysis by
Kartje et al. 1997; (c) during the remaining period, May 7-13 (MJD
49844-49851).   The best fit power law model (dotted line) and the
best fit power law + inverted gaussian model (solid line) are shown. 
The open circles 
in (b) correspond to the spectrum corrected with the on-axis effective
area (not accurate for this off-axis observation) as presented in
Kartje et al. 1997.\label{fig6}}

\figcaption{Same as in Fig.~2 for the three sections of the 1995
observation and the power law model.
They are displayed, from the top to the bottom, in temporal order:
part~1 corresponds to the flare, part~2 to the decay and part~3 
to the quiescent period.\label{fig7}}

\figcaption{EUV spectrum of MRK~421 obtained in 1996 with the SW spectrometer 
during a $0.3^{\circ}$ off-axis observation. The best fit power law
model (dotted line) and the best fit power law + inverted gaussian
model (solid line) are shown. \label{fig8}}
 
\figcaption{Same as in Fig.~2 for the 1996 spectrum and the power law
model \label{fig9}}

\figcaption{{\it ASCA} spectrum of MRK~421 obtained in 1995 during the flare.
The solid lines represent the best-fit model when an absorbed
power-law with a gradually changing spectral index is applied to the
spectrum of all the instruments simultaneously (upper panel ). The
lower panel shows the residuals.\label{fig10}}

\figcaption{EUV source flux  vs. energy index ($\alpha$)
for the power law component of the power law plus gaussian model.\label{fig11}}
\label{fig4}
\clearpage

\begin{deluxetable}{l l l l c c c}
\tablewidth{6.9in} 
\tablecaption{EUVE Observations of Mrk~421}
\tablehead{
\colhead{Year} & \colhead{Instrument\tablenotemark{a}} 
& \colhead {Start Date}
& \colhead {End Date}
& \colhead {T$_{\rm exp}$\tablenotemark{b}}
& \colhead {Count rate\tablenotemark{c}}
& \colhead {Spectral Data}\\
&&&& \colhead{(ks)}
& \colhead {(c s$^{-1}$)}
& 
}

\startdata
1994\tablenotemark{d}   &DS/S    &April 2       &April 12       &280    &0.18   &yes\\ 
1995            &Scanner &Feb. 4        &Feb. 7         &68     &0.45   &no\\ 
1995            &DS/S    &April 25      &May 13         &355    &0.29   &yes\\
1995 (flare) &DS/S    &April 25      &April 28       &86     &0.37 &yes\\
1995 (decay)\tablenotemark{e}  &DS/S    &April 29      &May 6          &154    &0.29&yes\\
1995 (quiescence)      &DS/S    &May 7         &May 13         &115    &0.22   &yes\\
1996            &DS/S    &April 17      &April 30       &299    &0.30   &yes\\ 
1996            &DS/S    &May 10        &May 11         &3.6    &0.30   &no\\ 
1997            &Scanner &Feb. 7        &Feb. 11        &108    &0.27   &no\\
1998 \tablenotemark{f}   &DS/S    &April 19      &May 1          &\nodata   &\nodata    &\nodata\\
\enddata

\tablenotetext{a} {``DS/S'' indicates that the source was observed simultaneously in the 3 
{\it EUVE} spectrometers and in the Deep
Survey photometer. ``Scanner'' indicates that the source was observed
simultaneously in the three {\it EUVE} photometers during a pointing
within the Right Angle Program.}
\tablenotetext{b}{Total exposure time calculated eliminating (i) all SAA
passages (ii) all satellite daytime data (iii) all times during which
the detector was turned off (iv) all times affected by possible earth
blockage. It does include corrections for telescope vignetting,
deadtime and limited telemetry allocation (primbshing).}
\tablenotetext{c}{Average count rate in the Lexan/B filter ($\approx 60-180$ \AA). 
The count rate in the scanner
has been normalized to the one in the DS assuming an average spectral
shape, to take into account the 
difference in the effective area between the two instruments.}
\tablenotetext{d}{This observation was presented in Fruscione et al. (1996}
\tablenotetext{e}{This part of the observation was presented in Kartje et al. 1997)}
\tablenotetext{f}{This observation is not included in this paper}
\end{deluxetable}

\clearpage

\begin{deluxetable}{l l c c c c}
\tablecaption{Spectral fits to the EUV spectrum of MRK~421 assuming an
absorbed power law model. Fixed parameters are $N_H=1.45 \times
10^{20}$~cm$^{-2}$, $N_{He{\sc I}}$/$N_{H{\sc I}}$=0.1  
and $N_{He{\sc II}}$/$N_{H{\sc I}}$=0.01}
\tablehead
{
\colhead{Obs.} 
& 
& \colhead{Range}
& \colhead{$\alpha$}
& \colhead{$10^{-3} f_{80 \AA}$\tablenotemark{a}}
& \colhead{$\chi ^2_{\nu}$}\\
& 
&\colhead{(\AA)}
&
&\colhead{(phot cm$^{-2}$ s$^{-1}$ \AA$^{-1}$)}
&
}
\startdata
1994    &               &75-110 &$0.85^{+0.66}_{-0.72}$ &$6.44 \pm 0.33$
&1.06\\
1995    &               &70-110 &$1.13 \pm 0.38$        &$6.70 \pm 0.22$
&2.45\\
{\it Flare}        &Apr. 25-28             &70-110 &$1.43 \pm 0.56$        &$10.28 \pm 0.51
$       &1.57\\
{\it Decay}        &Apr. 29-May 6  &70-110 &$2.19 \pm 0.50$        &$7.24 \pm 0.33$
&1.24\\
{\it Quiescence}        &May 7-13       &70-110 &$-0.60^{+1.13}_{-1.16}$ &$3.53 \pm 0.35$
&1.47\\
1996    &               &70-110 &$2.12 \pm 0.31$        &$9.32 \pm 0.27$
&1.61\\
\enddata
\end{deluxetable}

\clearpage

\begin{deluxetable}{l l c c c c c c c}
\rotate
\tablecolumns{9}
\tablewidth{9in}
\tablecaption{Spectral fits to the EUV spectrum  of MRK~421 assuming an
absorbed power law model plus a reverted gaussian. Fixed parameters as
in Table~2}
\tablehead
{
\colhead{}
&\colhead{}   
&\multicolumn{2}{c}{Power Law} 
&\multicolumn{3}{c}{Gaussian}
&\colhead{} 
&\colhead{} \\

\cline{3-4} \cline{5-7}\\
 
\colhead{Obs.}
&
& \colhead{$\alpha$}
& \colhead{f$_{80\rm{\AA}}$}
& \colhead{$\lambda_0$}
& \colhead{FWHM}
& \colhead{Ampl.}
& \colhead{$\chi^2_{\nu}$}
& \colhead{Flux (70-110 \AA)}\\

& 
& 
&\colhead{($10^{-3}\frac{\rm {phot}}{\rm{cm}^2 \rm{s} \rm{\AA}}$)}
&\colhead{(\AA)}
&\colhead{(\AA)}
&\colhead{($10^{-3}\frac{\rm {phot}}{\rm{cm}^2 \rm{s} \rm{\AA}}$)}
&
&\colhead{($10^{-3}\frac{\rm {phot}}{\rm{cm}^2 \rm{s}}$)}\\
}

\startdata

1994 & &$-0.29\pm0.83$ &$6.72\pm0.35$ &$79.3_{-0.07}^{+1.05}$
&$0.20\pm0.05$ &$-10.67\pm0.95$ &$1.14$ &5.06\tablenotemark{a}\\

1995 & &$1.14\pm0.41$ &$7.55\pm0.22$ &$73.5\pm0.84$  &$6.07\pm1.47$
&$-2.23\pm0.48$ &1.17 &8.86\\

Flare&Apr. 25-28 &$0.71\pm0.61$	&$11.6\pm0.52$ &$74.87\pm0.45$
&$3.04\pm0.95$ &$-5.86\pm0.14$ &0.57 &14.11\\

Decay&Apr. 29-May 6 &$1.62\pm  0.59$	&$7.84 \pm 0.34$ &$72.75\pm0.28$
&$1.86\pm0.51$ &$-5.22\pm1.30$&1.17 &9.33\\

Quiescence&May 7-13 &$0.63\pm1.29$	&$3.90 \pm 0.35$ &$71.43\pm0.04$
&$0.17_{-0.06}^{+0.23}$ &$-6.55\pm3.17 $ &1.29 &4.83\\

1996	& &$-0.15\pm 0.38$ &$9.64 \pm 0.28$&$71.82_{-0.60}^{+0.06}$
&$0.66\pm0.12$ 	&$-10.40\pm2.75$ &1.10 &12.02\\

\enddata
\tablenotetext{a}{This is calculated in the $75-110$ \AA\ range for the
model in Table 2}
\end{deluxetable}

\end{document}